\input{aipcheck}

\documentclass[
    ,final            
  ]
  {aipproc}

\layoutstyle{6x9}

\def\bge{\begin{equation}}
\def\ede{\end{equation}}

\begin{document}

\title{Galactic nuclei formation and activity induced by globular cluster 
merging}


\author{R. Capuzzo-Dolcetta}{
  address={Dep. of Physics, Univ. of Roma La Sapienza, P.le A. Moro, 00185, 
Roma, Italy}
}

\begin{abstract}
Different types of observations, together with consistent and physical modelizations,
suggest as realistic the hypothesis of
enrichement of 
galactic nuclei by mean of massive globular clusters orbitally decayed and 
merged in 
the inner regions of early type galaxies.
In this context, the scenario of globular cluster mergers and subsequent 
formation 
of a dense Super Star Cluster  in the center of a triaxial galaxy is presented
and discussed, together with its astrophysical implications, including that of  massive black
hole  feeding and accretion in the center of a triaxial galaxy.
\end{abstract}

\maketitle


\section{Introduction}
The Hubble Space Telescope and large ground based telescopes 
are providing an impressively increasing amount of data
concerning Globular Cluster Systems (GCSs) in galaxies, mainly of the early 
types,
since the pioneering work \cite{har79}.

Two are the most relevant and well defined observational points: (i) the 
difference in
the GCS and galaxy light spatial distribution,
and, (ii) the existence of a bimodal color distribution for GCSs,
and the possible differences between the {\it blue} and the
{\it red} population, 
\newline Here I will not discuss about point (ii)
(see the recent \cite{pen06} paper) but just about point
(i) whose solution implies an \lq evolutionary \rq  interpretation which has
relevant astrophysical implications.

\begin{figure*}
   \centering
   \resizebox{\hsize}{!}{\includegraphics[height=18cm]{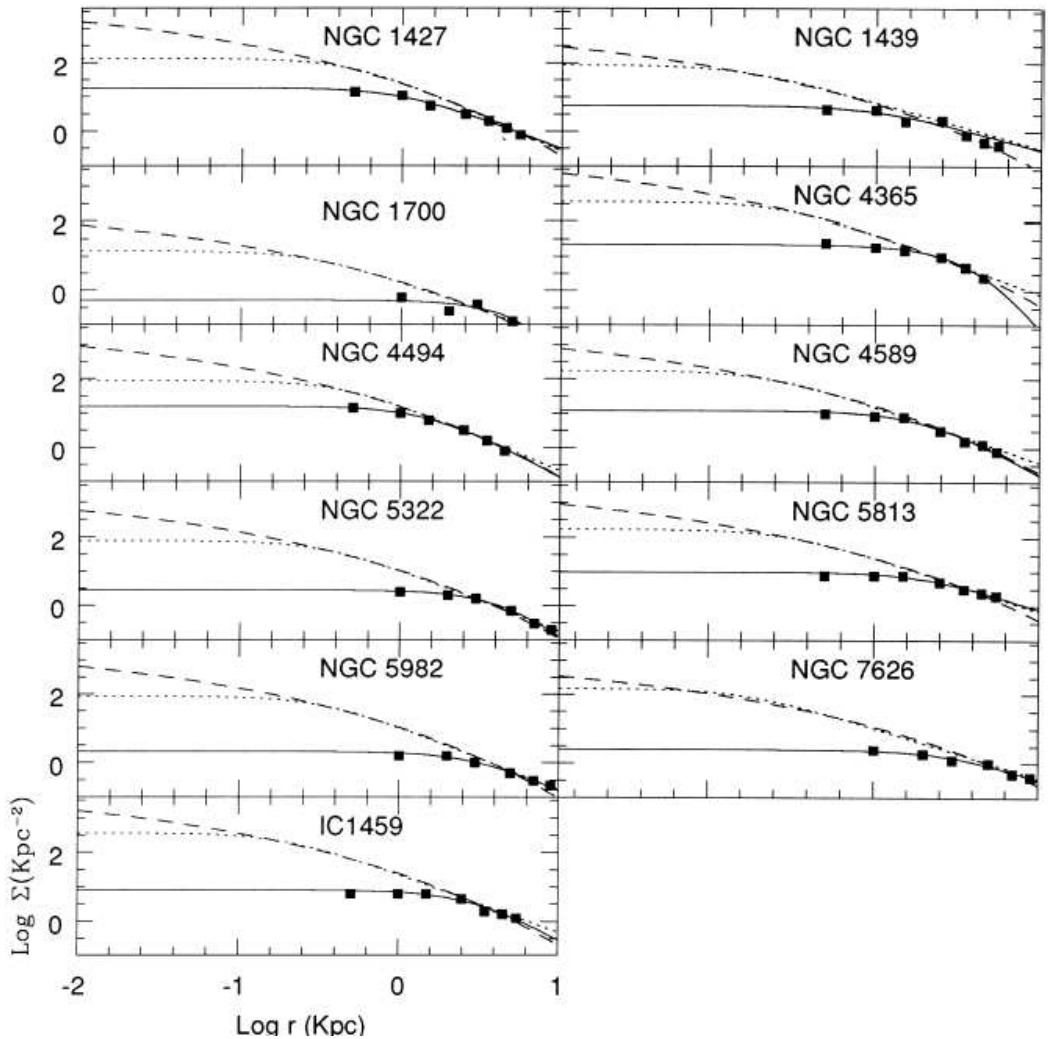}}
	\caption{Surface number densities for the galaxies of the \cite{for96} sample.
	Black squares represent the observed globular cluster distribution; the solid line
	is its modified core model fit. Dashed and dotted lines are de Vaucouleurs
	and modified core model fits to the normalized galaxy profile, respectively.
	The figure is taken from \cite{CDTes1999}.}
	\label{profiles}
\end{figure*}

\begin{figure}
  \includegraphics[height=.4\textheight]{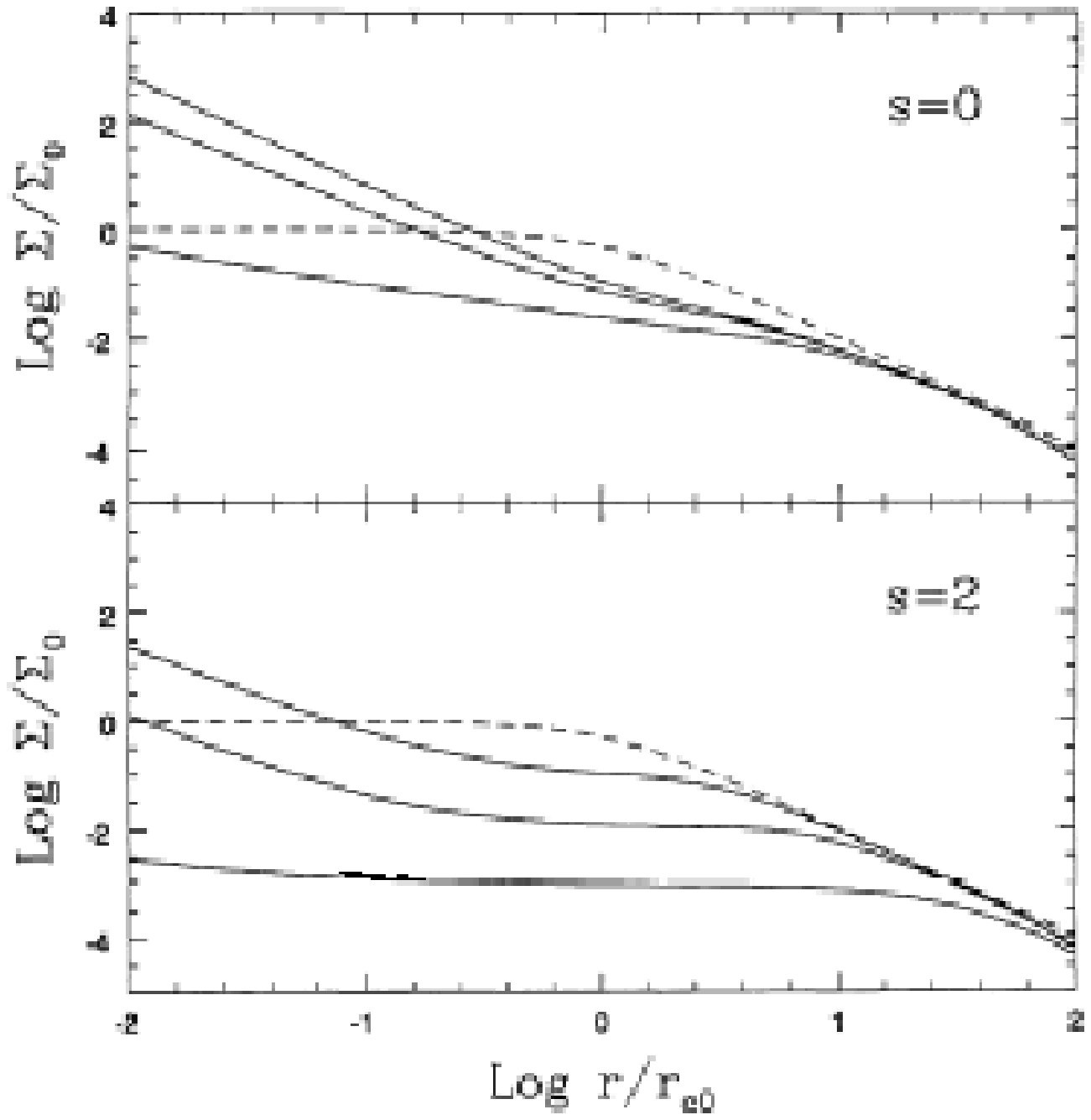}
  \caption{Plot of the GCS surface density profile after evolution has
  occurred for a flat (s=0) IMF of the GCS (upper panel) and for a steep (s=2) 
  IMF (lower panel).  The dashed curve is the initial profile; the other curves correspond to
  nuclear masses of $!0^7$, $10^8$ and $10^9$ M$_\odot$ (top to bottom).
  The figure is taken from \cite{CDTes1997}}
\label{theprofiles}
\end{figure}


\section{The GCS radial distributions in galaxies}
Presently available observations indicate clearly that the majority of galaxies
shows a radial profile of their GCS shallower than that of the stars toward
the galactic centre (see, for instance, Fig. \ref{profiles}).
Actually, ellipticals have usually a  stellar profile
peaked toward the galactic center (many have a \lq cuspy\rq~ profile, indeed),
while the GCS radial distribution shows, usually, a core. Among the many papers
about this topic, we limit to recall \cite{gri94},\cite{for96}. This difference
in the density profiles has an
interpretation either in terms of formation and evolution of elliptical galaxies
(see \cite{har86}, \cite{for96}, \cite{ash98}, \cite{bek06}).
or in terms of evolution of the GCS itself (see \cite{cap93},
\cite{CDTes1999}, \cite{cap05}).
\newline The explanation on the basis of GCS evolution is
more appealing, because much simpler and not based on qualitative and arbitrary
modelizations of GC formation in galaxies.
Moreover, it has other important astrophysical consequences, allowing to give an
answer to
the open question of the origin of the matter enriching massive black holes in
the center of active galaxies.
\subsection{GCS evolution}
The evolutionary view sketched above is remarkably simple because it bases just upon the, quite reasonable, hypothesis that the GCS and
the halo-bulge stars in the galaxy are
coeval and had, initially, the same radial density profile.
Under this assumption,  the presently observed different distributions
should be caused by evolution of the GCS, respect to the
unevolved, collisionless, halo-bulge stellar component.
\newline That GCSs in galaxies undergo an evolution is undoubtful,
because they are massive, evolving aggregates of stars moving in an
external potential which influences the system by both dynamical friction
and tidal distortion.
A detailed analysis of the GCS radial profile evolution in early type
galaxies under the combined influence of dynamical friction and tidal disruption,
mainly caused by a massive central black hole (bh), has been presented
in \cite{CDTes1999} where a quantitative
explanation of the observed comparative features of GCS' and stellar
light profiles is given. Fig. \ref{theprofiles} shows the
the expected projected GCS radial profile under the combined influence
of dynamical friction and tidal disruption (this latter mainly caused by a massive
central black hole), after an evolution of the GCS up to a Hubble time
\par
\begin{figure*}
   \centering
   \resizebox{\hsize}{!}{\includegraphics[height=18cm]{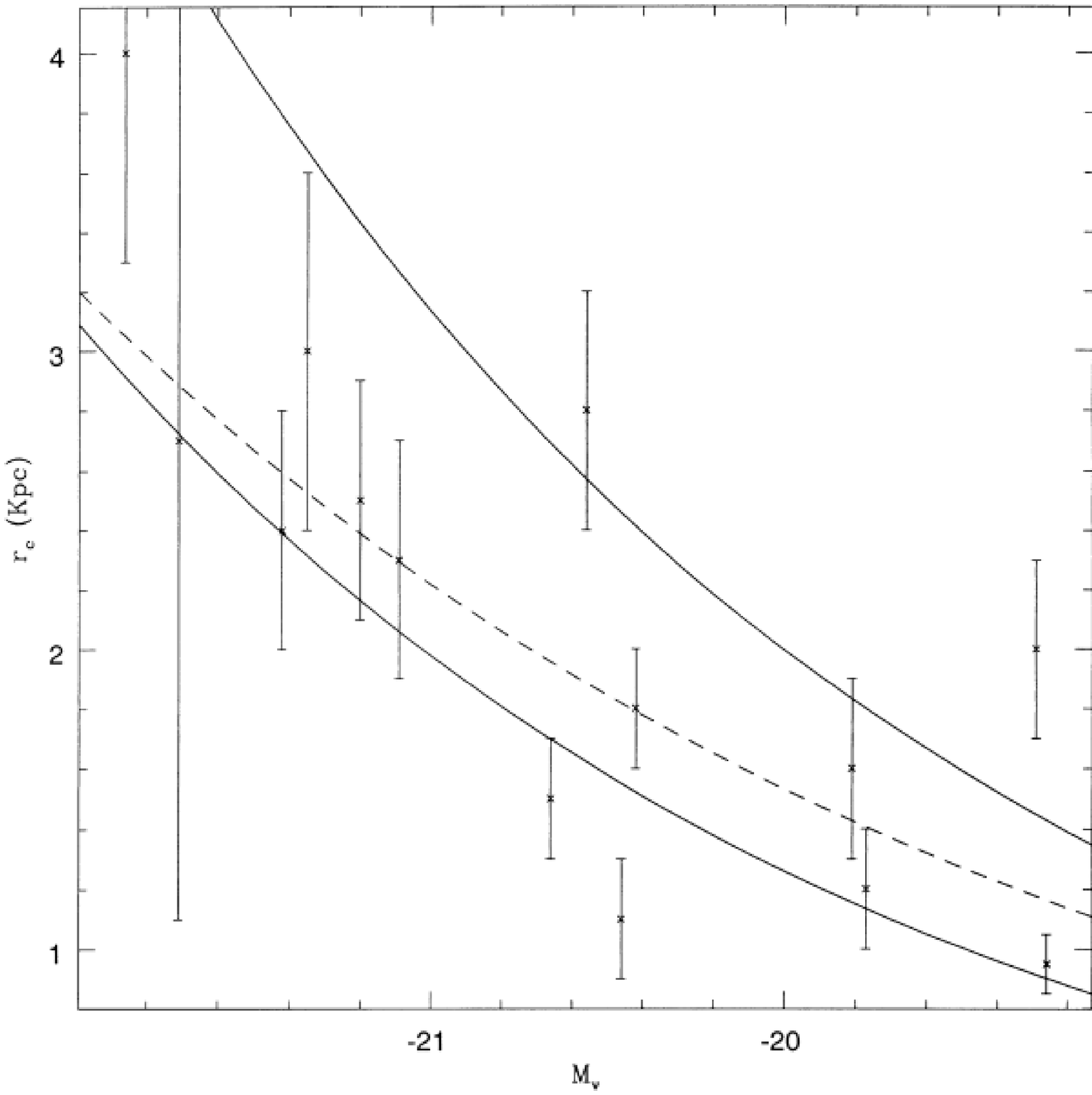}
   \includegraphics[height=17.5cm]{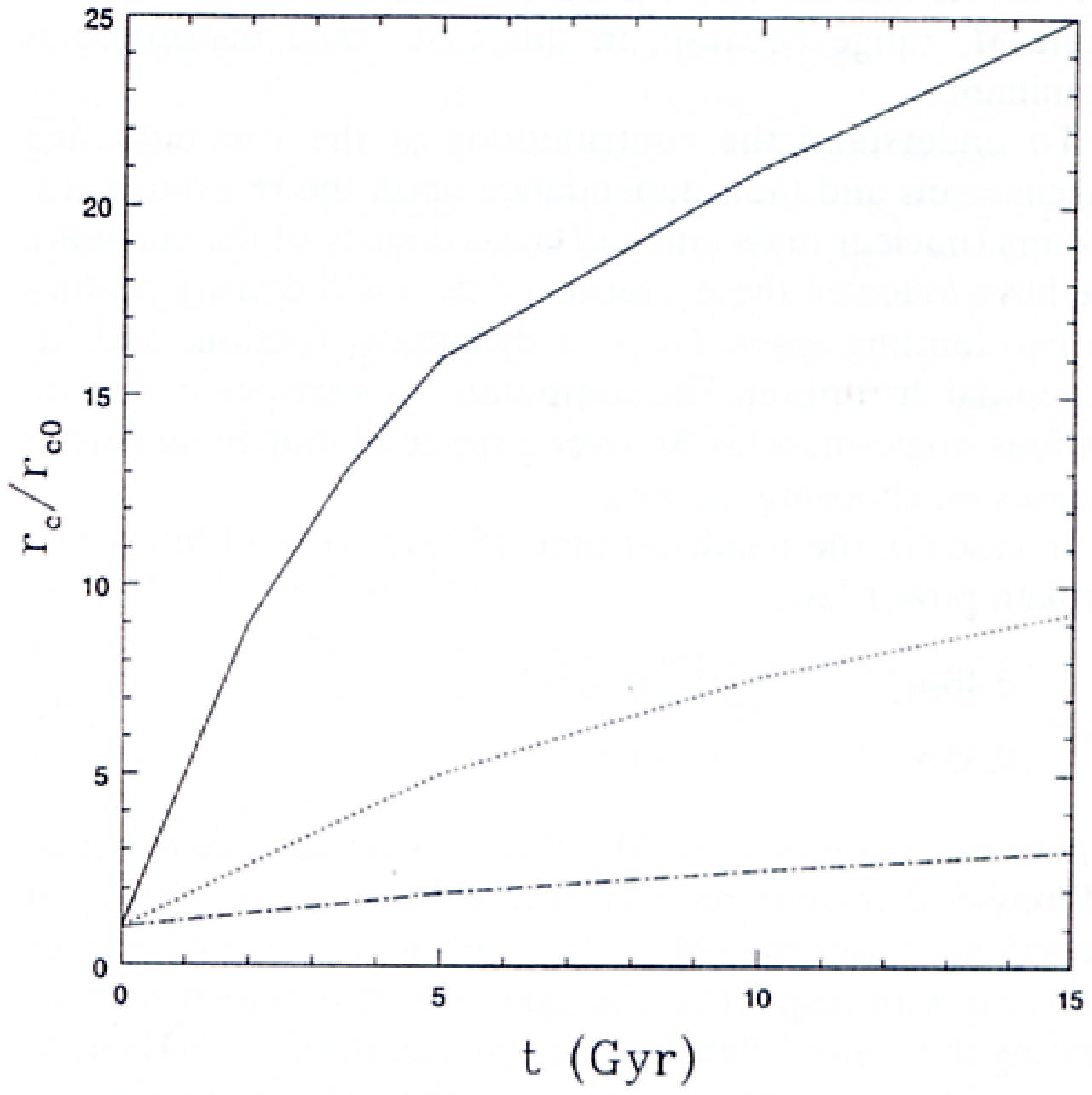}}
     \caption{Left panel: GCS core radius as a function of the absolute
integrated V mag. of the parent galaxy. Dots
     refer to data in \cite{for96}, with their best fit as dashed line. Solid
lines are two evoltionary models, with two different initial value of the GCS 
core radius (figure taken from \cite{CDTes1999}).
     Right panel: time evolution of the core radius of the GCS in a triaxial
galaxy containing a central bh of mass (from bottom up) $10^7$, $10^8$,
$10^9$ M$_\odot$ (figure taken from \cite{CDTes1997}).}
        \label{corerad}
    \end{figure*}

 Some researchers have invoked one observational
feature, the GCS radial distribution  being  shallower for brighter
galaxies than for faint (\cite{for96}), as evidence against
the \lq evolutionary\rq ~ explanation.
\newline Apart from that the claimed correlation is not universal
(for instance, \cite{bea06} found a quite shallow GCS radial
distribution in the Virgo dE VCC 1087), the evolution
of a GCS due to the combined role of dynamical friction, acting on the large scale, 
and nuclear tidal distortion, on both the large scale of the overall field
star distribution and on the smaller one of the compact galactic nucleus, 
leads to a positive correlation between the core radius of the GCS radial profile
and the galaxy integrated luminosity exactly as observed (see Fig.
~\ref{corerad}, left panel), which is clearly due to the increasing (with time)
GCS core radius size showed in the right panel of Fig. ~\ref{corerad}, which
depends on that the central galactic bh mass increases with time.
To conclude, the GCS slope vs. galaxy luminosity correlation is not,
unfortunately, a way to distinguish between the two above mentioned hypotheses
(compare left panel of Fig. ~\ref{corerad} with Fig. 4 in \cite{bek06}).
Anyway, it is relevant noting that the correlation found in \cite{bek06}
relies on ad-hoc assumptions, which the evolutionary scheme is free from.

\begin{table}
\label{dati}
\begin{tabular}{@{}lllllll}
\hline
Galaxy & $M_V$ & $M_{bh}$ & $N$ & $N_0$ & $\Delta N /N_0$ & $M_l$ ($M_{\odot}$) \\
\hline
Milky way & -20.60 & $2.60 \times 10^6$ & 155 & 211 & 0.27 & $1.80 \times
10^7$\\
M 31 & -19.82 & $2.30 \times 10^7$ & 283 & 368 & 0.23 & $2.30 \times 10^7$\\
M 49 & -23.10 & $5.00 \times 10^8$ & 6321 & 13080 & 0.52 & $2.23 \times
10^9$\\
M 87 & -22.38  & $3.61 \times 10^9$ & 4456 & 8021 & 0.44 & $2.33 \times 10^9$\\
NGC 1379 & -20.16 & - & 132 & 512 & 0.74 & $1.50 \times 10^8$\\
NGC 1399 & -21.71 & $5.22 \times 10^9$ & 5168 & 9680 & 0.63 & $1.44 \times
10^8$\\
NGC 1400 & -20.49 & - & 83  & 233 & 0.64 & $4.95 \times 10^7$\\
NGC 1404 & -20.49 & - & 508 & 1061 & 0.53 & $1.75 \times 10^8$\\
NGC 1407 & -21.77 & - & 317 & 407 & 0.22 & $3.40 \times 10^7$\\
NGC 1427 & -20.43 & $1.17 \times 10^8$ & 248 & 487 & 0.49 &  8.86  $\times
10^7$\\
NGC 1439 & -20.40 & $1.95 \times 10^8$ & 130 & 141 & 0.08 &  4.79 $\times
10^6$\\
NGC 1700 &  -21.52 & $4.37 \times 10^9$ & 25 & 39 & 0.36 &   3.66  $\times
10^6$\\
NGC 3258 & -21.40 & $1.00 \times 10^9$ & 305 & 512 & 0.40 & $6.80 \times
10^7$\\
NGC 3268 & -21.96 & - & 519 & 909 & 0.43 & $2.29 \times 10^8$\\
NGC 4365 & -22.06 & $7.08 \times 10^8$ & 517 & 849 & 0.39 &  $7.48 \times
10^7$\\
NGC 4374 & -22.62 & $1.00 \times 10^9$ & 4731 & 7177 & 0.34 & $8.20 \times
10^8$\\
NGC 4406 & -22.30 & $1.40 \times 10^8$ & 2834 & 4192 & 0.32 & $4.10 \times
10^8$\\
NGC 4494 & -20.94 & $4.79 \times 10^8$ & 200 & 297 & 0.33 &  2.98 $\times
10^7$\\
NGC 4589 & -21.14 & $3.09 \times 10^8$ & 241 & 371 & 0.35 &   7.58  $\times
10^7$\\
NGC 4636 & -21.71 & - & 1426 & 2149 & 0.34 & $1.55 \times 10^8$\\
NGC 5322 & -21.90 & $9.77 \times 10^8$ & 175 & 266 & 0.34 &   6.51  $\times
10^7$\\
NGC 5813 & -21.81 & $2.82 \times 10^8$ & 382 & 596 & 0.36 &  1.03 $\times
10^8$\\
NGC 5982 & -21.83 & $7.94 \times 10^8$ & 135 & 260 & 0.48 &   8.86  $\times
10^7$\\
NGC 7626 & -22.34 & $1.95 \times 10^9$ & 215 & 365 & 0.41 &   3.59 $\times
10^8$  \\
IC 1459  & -21.68 & $2.60 \times 10^8$ & 271 & 516 & 0.47 &  1.57 $\times
10^8$\\
\hline
\end{tabular}
\caption{The presently observed number of  clusters  ($N$), its initial value
($N_0$), its fractional variation ($\Delta N /N_0$), the mass lost in form of disappeared
globulars ($M_l$).
Data are from \cite{CDVig1997}, \cite{CDTes1999}, \cite{CDDon2001} and \cite{CDMas2006}.
The galaxy integrated magnitude ($M_V$) and central black hole mass ($M_{bh}$)
values are from the literature.}
\end{table}

\subsection{Mass loss from the GCS}
\begin{figure*}
   \centering
   \resizebox{\hsize}{!}{\includegraphics[clip=true]{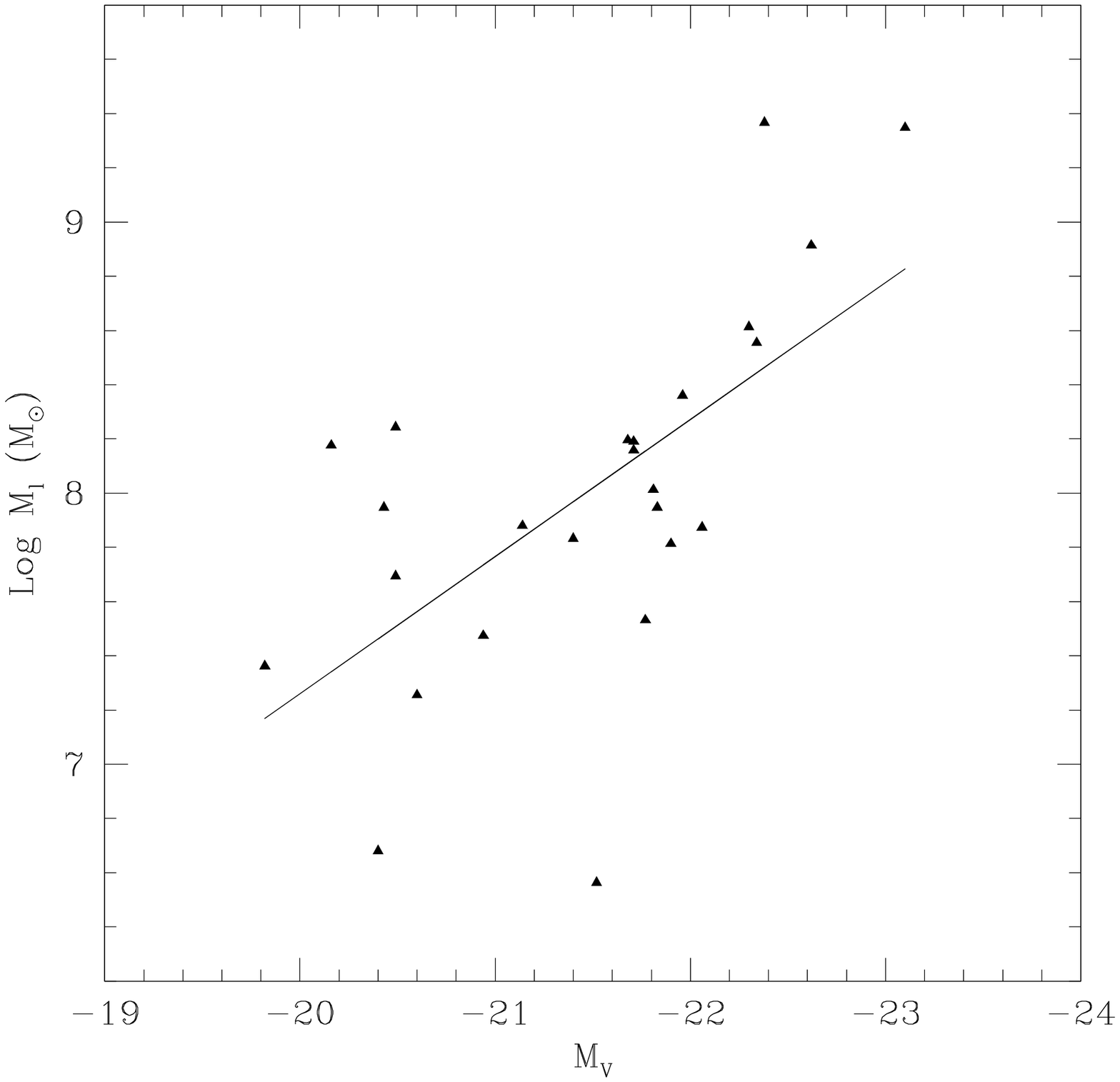}
   \includegraphics[clip=true]{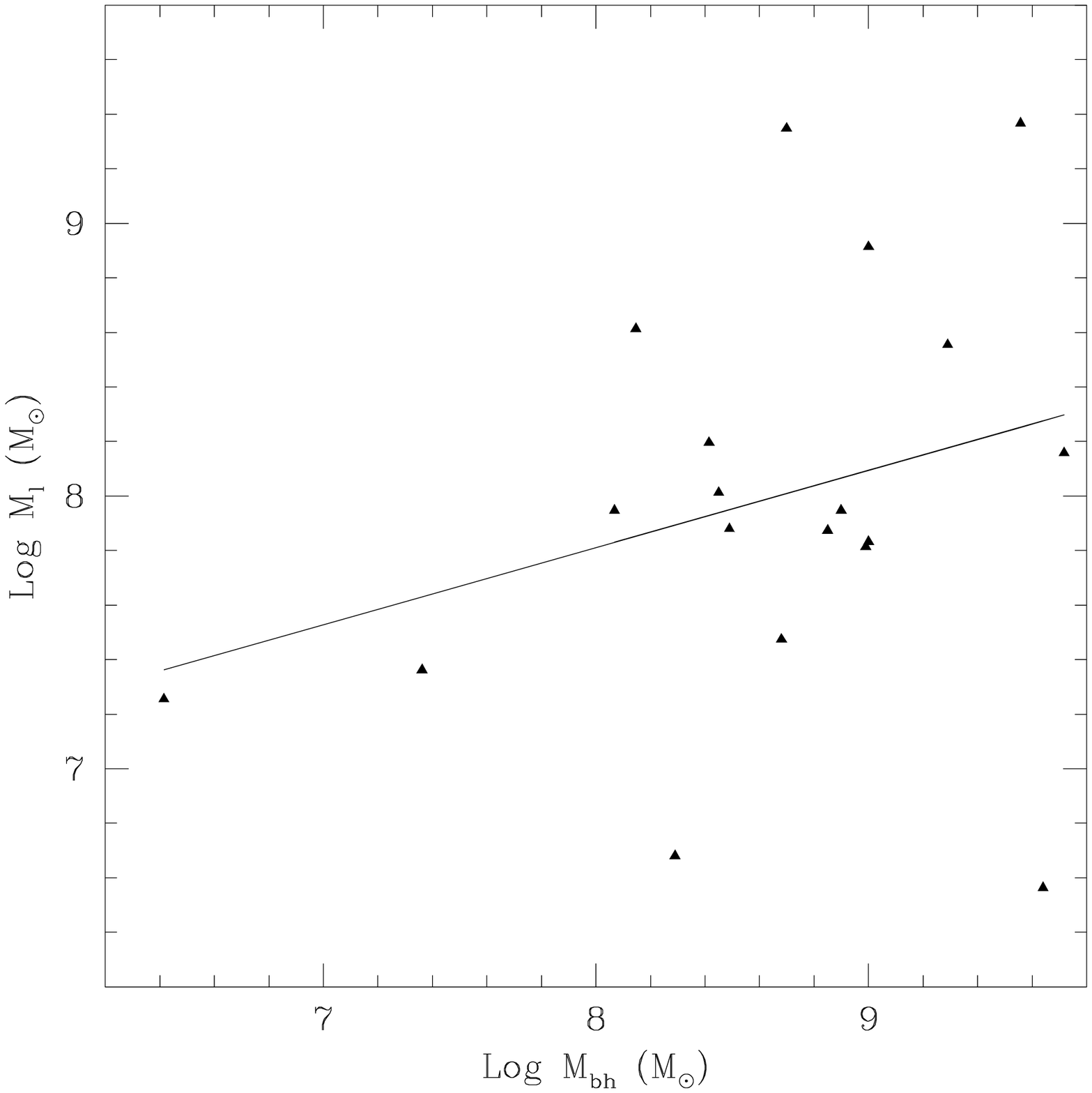}}
     \caption{Left panel: Mass lost from the GCS vs. the integrated galaxy
magnitude.
     Right panel: Mass lost from the GCS vs. the central black hole mass.}
        \label{mlost}
    \end{figure*}

Under the hypothesis that the flatter, respect to field stars,
central distribution of GCs is due to evolution and subsequent depauperation,
it is possible to evaluate the number of ~\lq lost \rq~ GCs by the difference of the
actually observed GCS radial profile and that of the bulge-halo stars,
which is considered as representative (after a linear scaling) of the GCS
initial distribution. This exercise was done for the first time by \cite{McL1995},
who estimated an upper limit on the total mass which could have been removed from
the M87 globular cluster system, yielding a value ($7.6\times 10^8$ M$_\odot$)
which is less than 30 \% of the size of the compact nucleus (supermassive black hole)
in M87. A much more detailed  study \citep{CDVig1997} of this
giant elliptical, as of the Milky Way and M 31, gave, instead, a much higher
value for the M87 GCS mass lost to the center, $M_l \approx 2.3 \times 10^9$ M$_\odot$,
i.e. $\sim 65 \%$ of the M87 bh mass. On this line, \cite{CDTes1999}, \cite{CDDon2001}
and \cite{CDMas2006} deduced values of the number and mass lost by GCSs
in several galaxies where good photometric data are available.
\newline Table \ref{dati} is an enlargement of what presented in
\cite{CDasp285} and \cite{CDDon2001}
to a set of 8 other recently studied GCSs (\citep{CDMas2006})
.
It resumes these old and new results on the number ($N_l$) and mass lost in form
of centrally decayed GCs, giving also the asbolute integrated magnitude of the
host galaxy and, when available, the central bh mass. The $M_V$ and $M_{bh}$ values
are collected from the literature and their discussion is postponed to a forthcoming
paper \citep{CDMas2006}. It is evident from data in the Table that the fraction (in number) 
of GCS eroded during a Hubble time is significant, ranging from $22\%$ of NGC 1407 to
$74\%$ of NGC 1379. Fig. \ref{mlost} shows the existence of a correlation between
$M_l$ and $M_V$ and between $M_l$ and $M_{bh}$. 
The least-square fits to the data are
\bge
LogM_l=-0.5059M_V-2.8592,
\ede
with rms$=0.5124$ and $\xi^2=6.5627$, and
\bge
LogM_l=0.2931LogM_{bh}+5.5461,
\ede
with rms$=0.2831$ and $\xi^2=5.5461$.
The large dispersion in Fig. \ref{mlost} is mainly due to inomogenheity of $M_V$ and $M_{bh}$
data, which deserves a more careful discussion.
This is quite interesting because it corresponds to a correlation of the GCS mass lost with 
both the {\it large} and the {\it small} scale structure of the galaxy. This is indeed
what expected in the frame of the evolutionary picture, as it is better explained
in the following.

\section{Super star cluster formation and nucleus accretion}
\label{ssc}
There is growing evidence of the presence of very massive young clusters,
up to the extremely large mass of W3 in NGC 7252 ($M = 8 \pm 2 \times 10^7$
M$_\odot$, \cite{mar04}). Massive clusters are, too, a significant fraction of galactic
GCs: actually, \cite{har06} indicates how up to a 40\% of the total mass
in the GCS of brightest cluster galaxies is contributed by massive GCs
(present day mass $> 1.5\times 10^6$ M$_\odot$), in good agreement with
recent theoretical results by \cite{kra05}.
\par The initial presence of massive clusters in a galaxy makes particularly
intriguing the GCS evolutionary frame sketched in the previous Sections, because
the presence of some massive primordial custers may have had very important
consequences on the initial evolution of the parent galaxy.
Actually, the GCS evolution in an elliptical galaxy naturally suggests the
following {\it scenario}:
\par (i) massive GCs on box orbits (in triaxial galaxies) or on low angular
momentum orbits (in axisymmetric galaxies) lose their orbital energy rather
quickly;
\par (ii) after $\sim 500$ Myr many GCs, sufficiently robust to tidal
deformation, are limited to move in the inner galactic region where
they merge and form a Super Star Cluster (SSC);
\par (iii) stars of the SSC buzz around the nucleus where some of them are
captured by a bh sitting there, partly increasing the BH mass;
\par (iv) part of the energy extracted from the SSC gravitational field goes
into electro magnetic
e.m. radiation inducing a high nuclear luminosity up to AGN levels.
\par Point (i) has been carefully studied in \cite{PCV} and \cite{cap05}
in self consistent models of triaxial core-galaxies, and presently under study
in triaxial
cuspy-galaxies with dark matter haloes \cite{caplecetal06}.
The validity of point
(ii) has been demonstrated by first results of \cite{cap93},
while the resistance to galactic tidal forces of sufficiently compact
GCs confirmed by \cite{miocchi06} and the actual formation of an SSC via
orbitally decayed cluster merger has been proved by detailed N-body
simulations \cite{capetal06}, \cite{mioetal06}.
Points (iii) and (iv) deserve a deeper investigation by mean of accurate
modeling, even if they seem reasonably well supported by previous studies
\cite{cap93},\cite{cap04}.

\section{Conclusions}
Various papers by our research group have shown that
many of the observed GCS features find a natural explanation in terms of
evolution of a GCS in the galactic field, assuming the (very conservative)
hypothesis that GCs were initially radially distributed as the galactic stellar
component and coeval to it. In other words, no {\it ad hoc} assumptions are needed to explain,
for instance,  the difference, observed in many galaxies, among the GCS-halo
star
profiles.
The initial presence of some massive GCs  ($M\geq 5\times 10^6$ M$_\odot$)
lead to the formation of a central SSC via merger of these orbitally
decayed massive clusters. The SSC mixed it up with the galactic nucleus in which
is embedded and constituted a mass reservoire to fuel and accrete a massive
object
therein.
Observationally, this latter picture is supported by the observed positive
correlation between the estimated  quantity of mass lost by a GCS in galaxies
and the mass of their central BHs (see Fig. 1 in \cite{cap01}).
On the theoretical side, the precise modes of mass
accretion onto the BH via star capture from the merged SSC still remain to be
carefully investigated.


\begin{theacknowledgments}
Thanks are due to P. Di Matteo, P. Miocchi, A. Tesseri, and A.
Vicari, whose collaboration has been precious in a significant part of the work
on the topics discussed here.
\end{theacknowledgments}



\bibliographystyle{aipproc}   

\bibliography{sample}

\IfFileExists{\jobname.bbl}{}
 {\typeout{}
  \typeout{******************************************}
  \typeout{** Please run "bibtex \jobname" to optain}
  \typeout{** the bibliography and then re-run LaTeX}
  \typeout{** twice to fix the references!}
  \typeout{******************************************}
  \typeout{}
 }


\end{document}